\documentclass[useAMS,usenatbib]{mn2e}
\usepackage{natbib}
\usepackage{amsmath,amsfonts}
\usepackage{epsfig}
\usepackage{mathptm}

\newcommand{\vect}[1]{{\mathbfit{#1}}}
\newcommand{\cosmomc}{{\sc cosmo-mc} }
\newcommand{\cog}{{\sc cog} }
\newcommand{\camb}{{\sc camb} }

\def\reff@jnl#1{{\rm#1\/}}

\def\aj{\reff@jnl{AJ}}                  
\def\araa{\reff@jnl{ARA\&A}}            
\def\apj{\reff@jnl{ApJ}}                
\def\apjl{\reff@jnl{ApJ}}               
\def\apjs{\reff@jnl{ApJS}}              
\def\ao{\reff@jnl{Appl.Optics}}         
\def\apss{\reff@jnl{Ap\&SS}}            
\def\aap{\reff@jnl{A\&A}}               
\def\aapr{\reff@jnl{A\&A~Rev.}}         
\def\aaps{\reff@jnl{A\&AS}}             
\def\azh{\reff@jnl{AZh}}                
\def\baas{\reff@jnl{BAAS}}              
\def\jrasc{\reff@jnl{JRASC}}            
\def\memras{\reff@jnl{MmRAS}}           
\def\mnras{\reff@jnl{MNRAS}}            
\def\pra{\reff@jnl{Phys.Rev.A}}         
\def\prb{\reff@jnl{Phys.Rev.B}}         
\def\prc{\reff@jnl{Phys.Rev.C}}         
\def\prd{\reff@jnl{Phys.Rev.D}}         
\def\prl{\reff@jnl{Phys.Rev.Lett}}      
\def\pasp{\reff@jnl{PASP}}              
\def\pasj{\reff@jnl{PASJ}}              
\def\qjras{\reff@jnl{QJRAS}}            
\def\skytel{\reff@jnl{S\&T}}            
\def\solphys{\reff@jnl{Solar~Phys.}}    
\def\sovast{\reff@jnl{Soviet~Ast.}}     
\def\ssr{\reff@jnl{Space~Sci.Rev.}}     
\def\zap{\reff@jnl{ZAp}}                
\def\nat{\reff@jnl{Nature}}             

\label{firstpage}
\title[An improved MCMC sampler for cosmology]{An improved Markov-chain 
Monte Carlo sampler for the estimation of 
cosmological parameters from CMB data}
\author[An\v{z}e Slosar and M.P.~Hobson]{An\v{z}e Slosar and M.P.~Hobson \\
Astrophysics Group, Cavendish Laboratory, Madingley Road, Cambridge
CB3 0HE}

\date{Accepted ---; received ---; in original form \today}
\pagerange{\pageref{firstpage}--\pageref{lastpage}}
\pubyear{2003}

\begin{document}

\maketitle

\begin{abstract}
Markov-chain Monte Carlo sampling has become a standard technique for
exploring the posterior distribution of cosmological parameters
constrained by observations of CMB anisotropies.  Given an infinite
amount of time, any MCMC sampler will eventually converge such that
its stationary distribution is the posterior of interest. In practice,
however, naive samplers require a considerable amount of time to
explore the posterior distribution fully.  In the best case, this
results only in wasted CPU time, but in the worse case can lead to
underestimated confidence limits on the values of cosmological
parameters. Even for the current CMB data set, the
sampler employed in the widely-used \cosmomc package does not sample
very efficiently. This difficulty is yet more pronounced for data
sets of the quality anticipated for the Planck mission. 
We thus
propose a new MCMC sampler for analysing total intensity 
CMB observations, which can
be easily incorporated into the \cosmomc software, but has rapid
convergence and produces reliable confidence limits.  This is achieved
by using dynamic widths for proposal distributions, dynamic covariance
matrix sampling, and a dedicated proposal distribution for moving
along well-known degeneracy directions.
\end{abstract}


\section{Introduction}
\label{sec:intro}

Markov-chain Monte Carlo (MCMC) sampling is a universal technique used
to explore high-dimensional density fields and has several advantages over 
more conventional grid-based approaches (see, for example, Gilks,
Richardson \& Speigelhalter 1995). In an astrophysical context, the
MCMC approach was first applied to the determination of cosmological
parameters from estimates of the total intensity 
cosmic microwave background (CMB)
power spectrum by Christensen et al. (2001) and Knox, Christensen \&
Skordis (2001). More recently Lewis \& Bridle (2002) released the
publicly-available \cosmomc software package for this purpose.
MCMC sampling methods have also recently been applied
to the detection and characterisation of Sunyaev-Zel'dovich clusters
in maps of primordial CMB anisotropies (Hobson \& McLachlan 2002).

In an MCMC algorithm, a Markov chain is constructed whose equilibrium
distribution is the density field (or posterior)
of interest $p(\vect{x})$. Thus, after propagating the Markov chain
for a given {\em burn-in} period, one obtains (correlated) samples
from $p(\vect{x})$, provided the Markov chain has converged.
A Markov chain is characterised by the fact that the 
state $\vect{x}_{n+1}$ 
is drawn from a distribution (or {\em transition kernel}) that
depends only on the previous state of the chain $\vect{x}_n$, and not
on any earlier state. 
In its simplest form an MCMC sampler can be constructed using the
Metropolis algorithm as follows. 
At each step $n$ in the chain, the next state $\vect{x}_{n+1}$ is chosen
by first sampling a {\em candidate} point $\vect{x}'$ from
some symmetric 
{\em proposal distribution} $q(\vect{x}|\vect{x}_n)=
q(\vect{x}_n|\vect{x})$, which may in
general depend on the current state of the chain $\vect{x}_n$. The
candidate point $\vect{x}'$ is then accepted with probability
\begin{equation}
\alpha = 
\begin{cases}
1 & \mbox{if $p(\vect{x}') > p(\vect{x})$}, \\
p(\vect{x})/p(\vect{x}') & \mbox{otherwise}. 
\end{cases}
\end{equation}
If the candidate point is accepted, the next state becomes 
$\vect{x}_{n+1}=\vect{x}'$, but if the candidate is rejected, the chain
does not move, so $\vect{x}_{n+1}=\vect{x}_n$. 
In the limit of infinite number of chain steps, the density
of the samples is proportional to the density $p(\vect{x})$, provided
the proposal function: (i) satisfies detailed balance, which requires that
the probability of proposing $\vect{x}'$  from $\vect{x}$ is the
same as probability to proposing $\vect{x}$ from $\vect{x}'$, and
(ii) is such that 
every point in the parameter space has a finite probability of
being proposed. 

Given an infinite number of steps any sampler satisfying the above
conditions will eventually reach a converged state in which the samples
are representative of target density $p(\vect{x})$.
In practice, however, the number of steps required
to achieve convergence and fully explore the target density
can vary dramatically depending on the form of
the proposal distribution. A good proposal 
function produces candidate points that have a high probability of
being accepted and ensures good mobility of the chain around the
target density.

A common choice for the proposal function, used for example by Knox et
 al. (2001), is a multivariate Gaussian centred on the current chain
position, with fixed widths $\sigma_i$ in each parameter
direction. There are, however, some problems associated with using
this proposal function alone as the single MCMC `engine'.  Firstly,
the proposal widths must be chosen with considerable care and tailored
to the application in hand.  If the proposal widths are set too large,
the acceptance rate will be very low, so that the chain remains stuck
at a single point for a considerable number of steps, sometimes
indefinitely in practical terms. On the other hand, if the proposal
widths are too narrow, the acceptance rate will be high, but the chain
has a limited mobility because it effectively executes a random
walk and thus diffuses only very slowly around the target density. In
particular, this can lead to severely underestimated confidence limits
on parameters and spurious peaks in the sample distribution associated
with the initial chain position.  A second problem with the standard
multivariate Gaussian proposal distribution is that, in
high-dimensional problems, the degeneracy directions often take the
form of small `tunnels' in the target density, which are unlikely to be
explored by chance. Even with appropriate proposal widths the chain
must `zig-zag' through such structures and therefore these degeneracy
directions are likely to be undersampled.

In an attempt to speed up
the sampling process, the \cosmomc package (Lewis \& Bridle 2002)
employs an number of devices that improve upon the simple multivariate
Gaussian proposal function; these are explained in Section~\ref{sec:cosmomc}
below. This widely-used package has proved very successful in
analysing CMB power spectrum measurements from ground-based and
balloon-bourne experiments.  Nevertheless, as we show in
Section~\ref{sec:application}, with the inclusion of WMAP data
(Bennett et al. 2003), which is cosmic variance 
limited out to $\ell \approx 350$,
the \cosmomc sampler still suffers from a number of the disadvantages 
listed above. In particular, it produces
marginalised probability distributions for some cosmological 
parameters that yield underestimates of confidence limits by 
up to a factor of two. This undersampling of the target density is 
yet more pronounced for cosmic variance limited CMB data out to
$\ell\approx 2000$, as is expected from the Planck mission.

In this paper, we therefore present a new
sampler module (called \cog) that is trivially substituted for the 
existing sampler in the \cosmomc software. 
This sampler employs a number of strategies to avoid the
difficulties encountered in the use of the standard multivariate
Gaussian proposal function and the native \cosmomc
sampler. These strategies are
explained in detail in Section~\ref{sec:cog}, but the single most important
advantage of the \cog sampler is its use of
the very well-known cosmological parameter degeneracies for CMB
data (Efstathiou \& Bond 1999), hence ensuring good mobility around the
target density. In Section~\ref{sec:application}, we test the \cog sampler
on the current CMB data set and a simulated data set of the quality
expected from the Planck mission.
In addition to producing reliable marginalised
distributions, the \cog sampler also requires far fewer evaluations
of theoretical $C_\ell$ spectra 
to explore the target density fully, and therefore
provides speed up by a factor $\sim 4$ over the standard \cosmomc software
when analysing the current CMB data set. In the analysis of the
Planck-like data set, this speed increase rises to a factor 
of $\sim 50$. Our conclusions are presented in Section~\ref{sec:conclusions}.

\section{The cosmo-mc sampler}
\label{sec:cosmomc}

The \cosmomc sampler uses 
\camb (Lewis, Challinor \&
Lasenby 2000) as its underlying theoretical
CMB power spectrum generator. The proposal density employed in this
single-chain sampler is based on a multivariate Gaussian, but is
tailored specifically to exploit the
difference between `fast' and `slow' parameters in {\sc camb}, in order to
increase the speed with which the target density may be sampled.

At any given point $\vect{x}_n$ in the chain, it is much
quicker to calculate the theoretical $C_\ell$ spectrum (and hence the
value of the posterior) at the next
candidate point $\vect{x}'$ if some of the parameter values
are the same for both points. In particular, since the perturbation evolution
is assumed linear, once the transfer function for each
wavenumber has been computed, it is fast to calculate the $C_\ell$
spectrum corresponding to changes in any parameters 
governing the initial primoridal power spectra of scalar and tensor
perturbations; these are thus termed `fast' parameters. On the other
hand, if one changes parameters governing the perturbation evolution,
the resulting $C_\ell$ spectrum will be much slower to compute since
it requires a detailed recalculation of the linear physics; these
are thus termed `slow' parameters.
 
In detail, the basic sampler works as follows. First it performs an
initial sampling of the target density with a proposal density based
on a multivariate Gaussian distribution determined by a set of user
defined proposal widths $\sigma_i$, one for each parameter. It begins
by placing the parameters in random order in a `queue'. If the first
parameter in the queue is fast, all the fast parameters are updated to
obtain the next candidate point $\vect{x}'$. If the first
parameter in the queue is slow, both this parameter and the next 0--2
parameters in the queue are updated (the number of additionally
updated parameters being drawn randomly from a uniform distribution). 
In either case, the corresponding candidate
point is then either accepted or rejected in the usual manner. In
subsequent iterations, the above process is repeated, each time
starting from the next parameter in the queue. Once the entire queue
has been looped over twice, the parameter order is again randomly
reshuffled, and the process repeated.  For each parameter that is
updated, the new parameter value is drawn from a Gaussian distribution
centred at the present parameter value and having width
$\sigma'_i=\sigma_i/\sqrt{N}$, where $N$ is the total number of
parameters changed. The extra $1/\sqrt{N}$ factor ensures that
proposals in which many parameters change still have a reasonable 
acceptance rate.

After the initial sampling stage, one then has the option of performing
a complete resampling of the the target distribution using information
gained from the inital stage. In particular, the empirical covariance matrix of
the initial set of samples is calculated, which is then 
diagonalised to obtain a
set of principal directions which are taken as the new `parameters'.
These new parameters are then placed in random order in a queue as
above, and at each proposal between 1 and 3 of them are updated (the
precise number of updated parameters again being chosen at random). 
Each time the queue has been looped through twice, it is again
randomly reshuffled. We note that a covariance matrix for the standard
6-parameter inflationary flat $\Lambda$CDM model is provided in advance in
the \cosmomc package. Thus, in this case, one can dispense with the
initial sampling stage altogether.

In spite of the increased sampling speed achieved by the above
devices, this basic sampler still suffers from the 
difficulties discussed in Section~\ref{sec:intro}
Most notably, the user-supplied 
proposal widths $\sigma_i$ must still be chosen with considerable
care to avoid acceptance rates that are either too low 
(so the chain becomes stuck) or too high (so the chain mobility is
limited, leading to underestimated confidence limits). Also, the use
of a multivariate Gaussian proposal function can lead to undersampling
along narrow degeneracy directions. 

\section{The cog sampler}
\label{sec:cog}

The \cog sampler is a replacement MCMC engine
for the \cosmomc software. It may be downloaded from 
\texttt{http://www.mrao.cam.ac.uk/$\sim$anze/cog} and included easily
into the \cosmomc package. It does, however, also require the 
Gnu Scientific Library  for the integration of the degeneracy 
quantities discussed below. 

The sampler is composed of four separate MCMC engines: fast
  parameter changes (E1), all parameter changes (E2),
  principal direction changes (E3), degeneracy 
direction changes (E4); each of these engines is explained below.
  At any given step in the chain, only one of engines is used to
  propose the next candidate point $\vect{x}'$. Which engine is
  used is decided randomly at each step according to a set of relative
  probabilities fixed by the user at compilation. To allow for the
  engines to respond dynamically to the stucture of the target
  density, at regular intervals (the number of chain steps defining an
  `interval' being fixed by the user at compliation: default 300) 
  the engines E1, E2 and E3
are overhauled. In particular, for E1 and E2, the proposal
  widths $\sigma_i$ are adjusted to ensure reasonable acceptance rates
  (see below). For E3, the empirical covariance matrix of recent
  samples is recalculated. We note that an overhaul is only performed after a
  rejection of a candidate point, thus ensuring that detailed balance
  is maintained.

\subsection{Fast parameter changes}

During a fast parameter change (engine E1), only fast parameters in
\camb are updated. Each fast parameter is assigned a probability
$p_i$ of being updated in any given E1 proposal. These are defined by
the user at compilation.  Note that the $p_i$ need not sum to
unity. If $N$ fast parameters are updated, each new parameter value is
drawn from a Gaussian distribution centred at the present parameter
value and having width $\sigma'_i=\sigma_i/\sqrt{N}$. For the first
sampling interval the $\sigma_i$ are those supplied by the user at
compilation. In contrast to
the \cosmomc sampler, however, the proposal width $\sigma_i$ for
each parameter is then updated at each overhaul, based on the acceptance
rates achieved for that parameter in the previous sampling
interval. If the average acceptance rate of a given parameter is less
than a user selected target $\beta$ (defined at compilation: default 0.4), 
it is likely that the proposal width for this parameter is
too large and so the corresponding $\sigma_i$ is decreased by a fixed
factor (defined at compilation: default 0.8).  Similarly,
if the acceptance rate is less than $\beta$ the proposal width is probably too
narrow and is thus increased by a fixed factor 
(defined at compilation: default 1.2). As we show in
Section~\ref{sec:application},
the proposal width for each parameter eventually settles down to a
stable value appropriate to the target density being sampled.

\subsection{All parameter changes}

The all-parameter-change engine (E2) operates in an identical manner
to E1, except that in this case the full set of fast and slow 
parameters may be updated.

\subsection{Principal-direction changes}

The engine (E3) implements prinicipal-direction changes and makes use
of the covariance information collected from early samples. This engine
is only switched-on once the chain has taken a given number of steps
(defined at compilation: default 200), at which point the empiricial
covariance matrix of the parameters is calculated using 
preceeding samples; an upper limit on the number of preceeding samples
used in the calculation is defined at compilation (default 2000).
The covariance matrix is then updated at each overhaul.
The eigenvectors and eigenvalues of the covariance matrix are then
determined, which yield the principal directions in which changes 
may be proposed and the corresponding Gaussian proposal widths. In an analogous
manner to engines E1
and E2, each principal direction is assigned a probability $p_i$ of
being updated (defined at compilation). Once again the $p_i$ need not
sum to unity. 

This approach has two advantages over the covariance matrix method used
in the \cosmomc sampler. Firstly, the
covariance matrix corresponds to the local degeneracy directions and
is thus optimised for the current position of the chain. Secondly, it
alleviates the need to run the entire sampling process twice, thus making the
computation requirements considerably smaller.

\subsection{Degeneracy-direction changes}

The degeneracy directions engine (E3) is the most important innovation
in the \cog sampler, and ensures good mobility of the chain
around the target density. This engine takes advantage of the fact that
there are degeneracies in cosmological parameter space
that are intrinsic to any CMB observation and cannot
be broken even with cosmic-variance limited data. 
Some of these degeneracies have been extensively analysed in Efstathiou and 
Bond (1999). In particular, we implement motion of the chain
along the two most important degeneracies: namely the geometrical
degeneracy and the peak position degeneracy.
In addition, we also implement motion of the chain along
the degeneracy directions spanned by the parameters 
$(z_{\rm re},A_{\rm s})$ and $(\omega_{\rm b},n_{\rm s})$.
The probability
$p_i$ of updating along each degeneracy direction is defined at
compilation (default 0.25).

\subsubsection{Geometrical degeneracy} 

The geometrical degeneracy is a nearly exact degeneracy in 
cosmological parameter space for CMB data; it therefore allows
large movement of the chain in the space. According to
Efstathiou and Bond (1999), two cosmological models
are degenerate if they have the same physical matter
densities $\omega_{\rm b}$ and $\omega_{\rm dm}$, 
the same primordial scalar and tensor
fluctuation spectra and the same value of parameter
\begin{equation}
{\cal R} = \omega_m^{1/2} y S\left(\omega_k^{1/2} y \right), \label{eq:es1}
\end{equation}
where $\omega_{\rm m}=\omega_{\rm b}+\omega_{\rm dm}$, $y$ is given by
the integral
\begin{equation}
y  = \int_{a_r}^{1} {{\rm d}a \over [\omega_m a + \omega_k a^2 +
\omega_\Lambda a^4]}, \label{eq:es2}
\end{equation}
and
\begin{equation}
S(x) = \frac{\sinh (x)}{x}.
\end{equation}
For a flat universe the function $S$ equals unity, 
and for closed universes it simplifies to ${\rm sinc}(\sqrt{-\omega_k}  y)$. 
The integral (\ref{eq:es2}) may be evaluated
numerically, where the lower limit is obtained using the fitting
formula for the redshift of recombination  given by Hu and Sugiyama
(1995), namely
\begin{eqnarray}
z_r & = &1048[ 1 + 0.00124 \omega_b^{-0.738} ]
[ 1 + g_1 \omega_m^{g_2}]\, , \label{eq:20}\\
g_1 & = &  0.0783 \omega_b^{-0.238}
[ 1 + 39.5 \omega_b^{0.763} ]^{-1}\, , \nonumber \\
g_2 &=&  0.560 [ 1 + 21.1 \omega_b^{1.81} ]^{-1} \, . \nonumber
\end{eqnarray}
The quantity ${\cal R}$ is thus a function of the parameter set
$(\omega_{\rm b}, \omega_{\rm m}, \omega_\Lambda, \omega_{\rm k})$. 
These parameters can be
calculated from the equivalent \cosmomc parametrisation 
$(\omega_b, \omega_{\rm dm}, \Omega_k, h)$.

When a geometric degeneracy change is proposed, the parameter ${\cal R}$ is
calculated for the current chain position $\vect{x}$. 
A fixed-width Gaussian change in
$\omega_\Lambda$ is proposed and the $\omega_k$ direction is then
searched using a numerical minimiser until a model with a matching
${\cal R}$ is found. Finally, the new set of parameters are converted back to
the standard \cosmomc parametrisation for this new candidate point
$\vect{x}'$. The proposal width in $\omega_\Lambda$ is
set at compilation and can be quite large (default $0.1$), 
since the degeneracy is almost exact so candidate points should have a
high acceptance rate. Indeed, the difference in CMB power spectrum
between models $\vect{x}$ and $\vect{x}'$ is below the
level of cosmic variance. However, due to inaccuracies in
above approximations, as well as intrinsic numerical 
inaccuracies in the \camb code, the
acceptance rate for these changes is less than unity.
Finally, we note that although the
numerical minimisation step involves many numerical integrations, the time
spent searching this space is negligible compared to the calculation
of a single CMB power spectrum.  

\subsubsection{Peak position degeneracy} 

Another important
degeneracy is the first peak position degeneracy. Although not as exact as
geometrical degeneracy it still allows the chain to make large steps in the
parameter space. Following Efstathiou \& Bond (1999), the position of the
first peak is approximately given by
\begin{equation}
\ell_d \approx 0.746 \pi \sqrt{3} a_{\rm r}^{-1/2}
\frac{\cal R}{I_s(\omega_m,\omega_b) }
\end{equation}
where
\begin{equation}
I_s = a_{\rm r}^{-1/2}
\int_0^{a_r} \frac{{\rm d}a}{\sqrt{(a + a_{\rm eq})(1+R)}}
\end{equation}
in which $a_{\rm eq}$ is the normalised scale factor at matter-radiation
equality and $R = (3 \rho_b/4 \rho_\gamma)$. To a good approximation
\begin{eqnarray}
a_{\rm eq}^{-1} & = & 24185\left ({1.6813 \over 1+\eta_\nu} \right )\omega_m,\\
\qquad R & = & 30496 \omega_b a\, . \label{eq:18}
\end{eqnarray}
Thus $\ell_d$ is also a function of the parameter set 
$(\omega_b, \omega_m, \omega_\Lambda,
\omega_k)$. Since $\omega_b$ and $\omega_m$ are not required to
be fixed there is considerably more freedom in choosing which
parameters to change.  In the present implementation $\omega_m$ is always
changed with $\omega_b$ and $\omega_k$ being changed with some finite
probability. The $\omega_\Lambda$ direction is then numerically
searched for a model with a matching first peak position.

We note in passing that the third degeneracy mentioned in Efstathiou
\& Bond (1999), i.e. the height of the
first peak degeneracy, is not very relevant to future analysis, since
it is already broken by the existing CMB dataset.

\subsubsection{The $z_{\rm re}$--$A_{\rm s}$ degeneracy}

Another well-known approximate physical degeneracy in the cosmological
parameter space for CMB data is that between $\sigma_8$ and
$\exp(-\tau)$ (see e.g. Lewis \& Bridle 2002) This occurs because the
CMB power spectrum on scales smaller than the horizon size at
reionisation is damped by a factor $\exp(-2\tau)$, and $\sigma_8^2$
scales with the power in the primordial perturbation.  The
corresponding parameters in the \cosmomc software are $z_{\rm re}$ and
$A_{\rm s}$. We note that $z_{\rm re}$ is a `slow' parameter, whereas
$A_{\rm s}$ is a `fast' one. When a move in $z_{\rm re}$ is proposed,
the sampler additionally proposes up to 30 (or a number selected at
the compilation time) proposals in $A_{\rm s}$. The proposal
probability distribution function for $A_{\rm s}$ is a half Gaussian
with the correct orientation (i.e. increasing $z_{\rm re}$ requires
increasing $A_{\rm s}$ and therefore sampler always proposes change
which has the right orientation with respect to the degeneracy
direction). Since $A_{\rm s}$ is a fast parameter the computational
cost is essentially just one \camb call.

\subsubsection{The $\omega_{\rm b}$--$n_{\rm s}$ degeneracy}

There is a weak degeneracy in the parameters $\omega_{\rm b}$--$n_{\rm
 s}$. We have implemented a degeneracy sampling system that is
 analogous to that described in the previous section with the slow
 parameter $w_{\rm b}$ and the fast parameter $n_{\rm s}$.

\subsection{Burn-in, convergence and annealing}

As mentioned in Section~\ref{sec:intro}, any MCMC sampler requires a burn-in
period for the chain to reach equilibrium and hence sample from the
target distribution $p(\vect{x})$. 
Unfortunately, there exists no
formula for determining the length of the burn-in period, or for
confirming that a chain has reached equilibrium. Indeed, the topic
of convergence is still a matter of ongoing statistical research.
Nevertheless, several {\em convergence diagnostics} for determining
the length of burn-in have been proposed. These employ a variety of
theoretical methods and approximations that make use of the output
samples from the Markov chain. A review of such diagnostics is given
by Cowles \& Carlin (1994). It is worth noting that
running several parallel chains, rather than a single long chain, 
can aid the diagnosis of convergence, as discussed below in
Section~\ref{sec:application}. 

It is also useful during burn-in to employ
an {\em annealing schedule} in which the target density is raised
to some power $\lambda$, which varies gradually from zero to unity.
Thus, one begins
sampling from a modified posterior with $\lambda=0$ and 
slowly raises $\lambda$ according to some (geometric)
{\em annealing schedule} until $\lambda=1$. This allows
the chain to sample from remote regions of the posterior
distribution, which in turn facilitates extensive chain mobility and ensures
more reliable relaxation of the chain into the global optimum.
This approach also has the convenient by-product of
yielding an estimate from the burn-in samples 
of the Bayesian evidence for the model under 
consideration (see, for example, Hobson \&
McLachlan 2002; Slosar et al. 2003).

\section{Application to CMB data}
\label{sec:application}

To test the quality of our improved sampler, we have performed 
parameter estimation in a 7-parameter inflationary $\Lambda$CDM model 
the using the two separate data sets outlined below. 
We have
compared the results from the \cog sampler with those obtained
using the standard \cosmomc software. For both samplers, the
initial starting position of the chain and the initial proposal 
widths $\sigma_i$ for each parameter were identical, and are listed in 
Table~\ref{t1}. Also listed are the limits of the
top-hat priors adopted for the parameters. Both the \cosmomc and \cog
samplers were run until 6000 post burn-in 
accepted samples had been collected for
each chain. For both samplers, eight independent chains were run on
separate nodes of a Beowulf cluster. Thus, the total number of 
post burn-in 
accepted samples was 48000. This approach also allows one to compare 
samples obtained from different individual chains, which aids the
determination of convergence.
\begin{table}
\begin{center}
\begin{tabular}{lccc}
\hline
parameter & $\vect{x}_0$ & top-hat prior & $\sigma_i$ \\
\hline
$\omega_{\rm b}$ & 0.023 & $(0.005,0.1)$ & 0.01 \\
$\omega_{\rm dm}$ & 0.127 &  $(0.01,0.9)$ & 0.1 \\
$h$ & 68 & $(40,100)$ & 15 \\
$z_{\rm re}$ & 13 & $(6,20)$ & 6 \\
$\Omega_{\rm k}$ & 0.0 & $(-0.3,0.3)$ & 0.06 \\
$n_{\rm s}$ &  1.0  &  $(0.7,1.3)$ & 0.1 \\
$A_{\rm s}$ & 23 & $(10,50)$ &  15.0 \\
\hline
\end{tabular}
\caption {The initial chain position $\vect{x}_0$, the range of
  top-hat prior, and the initial proposal width $\sigma_i$ used for each
  parameter. The $\sigma_i$ were chosen to be small to give the original
\cosmomc sampler some chance of exploring the cosmic variance
  limited model.\label{t1}}
\end{center}
\end{table}

\subsection{Current CMB data}

The first data set (data set 1)
consists of WMAP, VSA, ACBAR and CBI band-power measurements of the 
total intensity CMB power spectrum (Bennett et al. 2003; Grainge et
al. 2003; Kuo et al. 2003; Pearson et al. 2003); for $\ell < 800$ only
points from the WMAP experiment are used. In the calculation of the
WMAP likelihood we use an adapted version of the publicly-available 
code (Kogut et al. 200; Hinshaw et al. 2003; Verde et al. 2003).

\subsubsection{Acceptance rates and sampler performance}

\begin{figure}
\epsfig{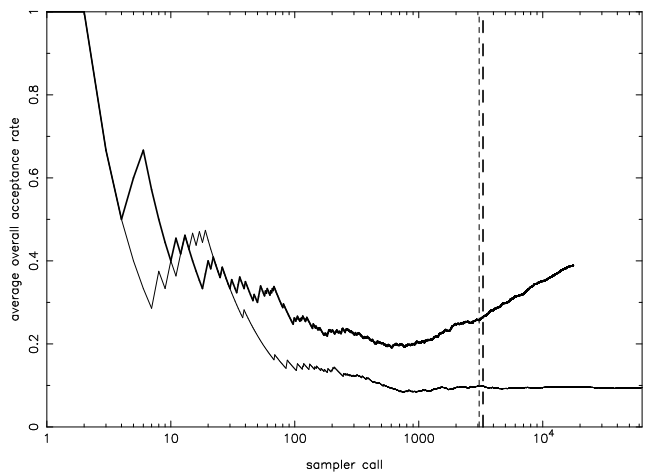}
\caption{The average acceptance rate as a function
of the number of sampler calls 
for the \cosmomc sampler (thin line) and the \cog
sampler (thick line) in the analysis of data set 1.
The vertical lines indicate the end of the burn-in period for
the \cosmomc (thin dashed line) and \cog (thick dashed line) samplers.
\label{d1fig0}}
\end{figure}

The average acceptance rates for the \cosmomc and \cog samplers
are plotted in Fig.~\ref{d1fig0} as a function of the number of calls
to the sampler.
The original \cosmomc sampler was supplied
with the precomputed basic 6-parameter covariance matrix that comes
supplied with the \cosmomc software; this includes all the parameters
given in Table~\ref{t1} except for $\Omega_{\rm k}$. For this last
parameter, \cosmomc uses a Gaussian proposal distribution with the
width indicated in the table. The provision of the covariance matrix
in advance gives the \cosmomc sampler
some advantage as compared with its own `initial' sampling phase.
Nevertheless, the \cog sampler at first equals this
acceptance rate, and very soon surpasses it, once it has learnt 
the appropriate length scales of the 
posterior. After many samples, the acceptance rate for the \cosmomc
sampler is found to be around 0.1. To obtain 6000 post burn-in 
accepted samples,
65000 calls to the sampler were made.
By contrast, the average acceptance rate of the \cog
sampler is around 0.4 and only 17000 calls to sampler were required.
\begin{figure}
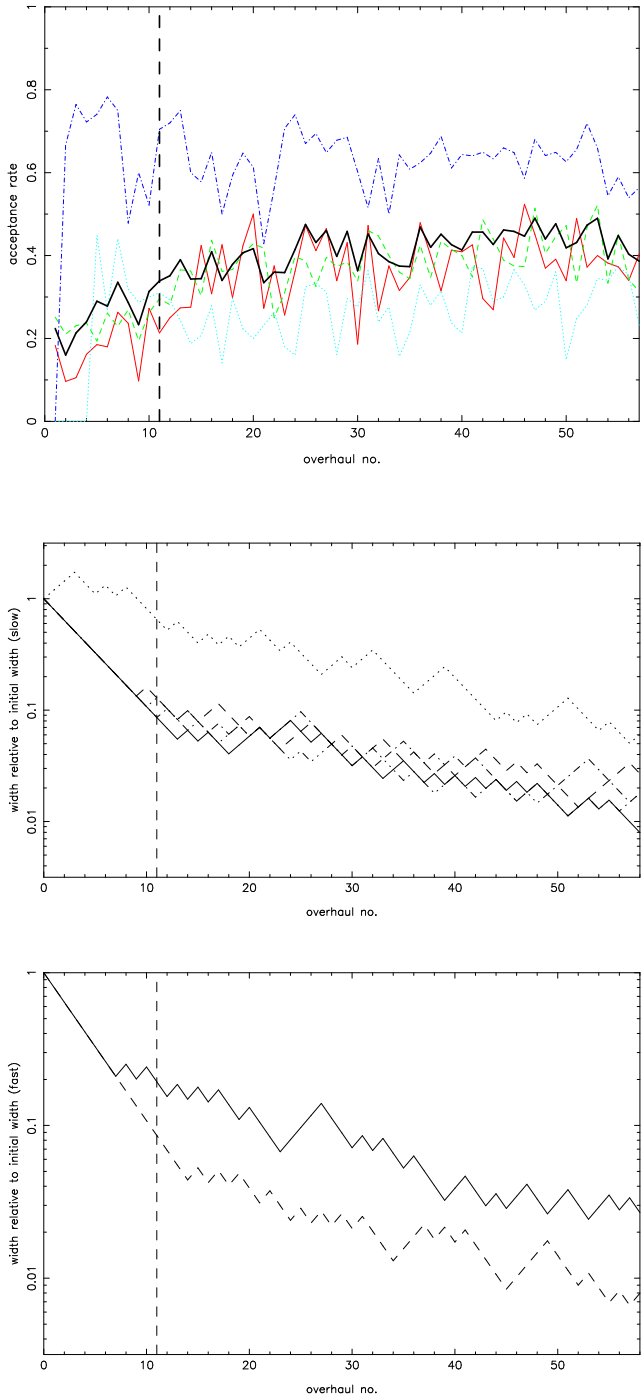

\epsfig{file=acc1.ps,height=\linewidth, angle=-90}\\[10mm]
\epsfig{file=wid1.ps,width=\linewidth}
\caption {Top: the acceptance rates in the analysis of data set 1
for each of the four MCMC engines
used in the \cog sampler as a function of the number of overhauls,
which are performed after every 300 chain steps -- fast 
parameter (E1: thin solid line), all parameters (E2: dashed line), principal
directions (E3: dotted line), degeneracy directions (E4: dot-dashed line).
The overall acceptance rate for the sampler is shown as the thick
solid line. Middle: the proposal widths $\sigma_i$ for 
each slow parameter as a function of overhaul number. 
Each width is expressed as a
fraction of its initial value given in Table~\ref{t1}, 
for each slow parameter as a function of overhaul number. 
Bottom: as above, but for the fast parameters.\label{d1fig1}}
\end{figure}

In Fig.~\ref{d1fig1} (top panel), 
we plot the acceptance rates for each of the four MCMC engines
used in the \cog sampler as a function of the number of overhauls,
which are performed after every 300 chain steps. As one might expect,
the engines E1 and E2 have acceptance rates of around 0.4, since the
proposal widths for the parameters are chosen to achieve this target.
More interestingly, the E3 engine also has an acceptance
rate of almost 0.4, once equilibrium has been achieved. We note, however,
that the initial acceptance rate for E3 is
zero, since there are not enough samples to compute the covariance
matrix. For E4 (degeneracy direction changes), one might have expected
the acceptance rate to be high, since the change in the
theoretical CMB power spectra corresponding to such a proposal 
is generally quite small. Nevertheless, inexactness in the degeneracies
and small numerical inaccuracies in both in \camb and the calculation
of the degeneracy parameters means that the acceptance rate is reduced
somewhat below unity, although its equilibrium value of around 0.6 is
higher than for the other engines. Moreover, successful proposals along the
degeneracy directions take the chain a considerable distance from its
original position and are thus very important.

In Fig.~\ref{d1fig1} (middle and bottom panels), we plot the
proposal widths $\sigma_i$ for the slow and fast parameters respectively,
as a function of overhaul number; each width is expressed
as a fraction of its initial value as given in Table~\ref{t1}. 
We see that each proposal widths initially decrease and then start oscillating
around their equilibrium values. This oscillation could be avoided by
adding a damping term to the proposal width change function, but we
believe that oscillating gives the sampler opportunity to sample a larger
range of distances and potentially escape local minima while at the
same time maintaining high acceptance rate. Additionally we note that the
requirement to maintain an acceptance rate of $0.4$ means that the proposal
widths are considerably smaller than widths of inferred uncertainties
in parameters. We have performed simulations on a multivariate
Gaussian and results show that this is indeed an effective
sampling technique. 

\subsubsection{Inferred limits on cosmological parameters}

\begin{figure}
\epsfig{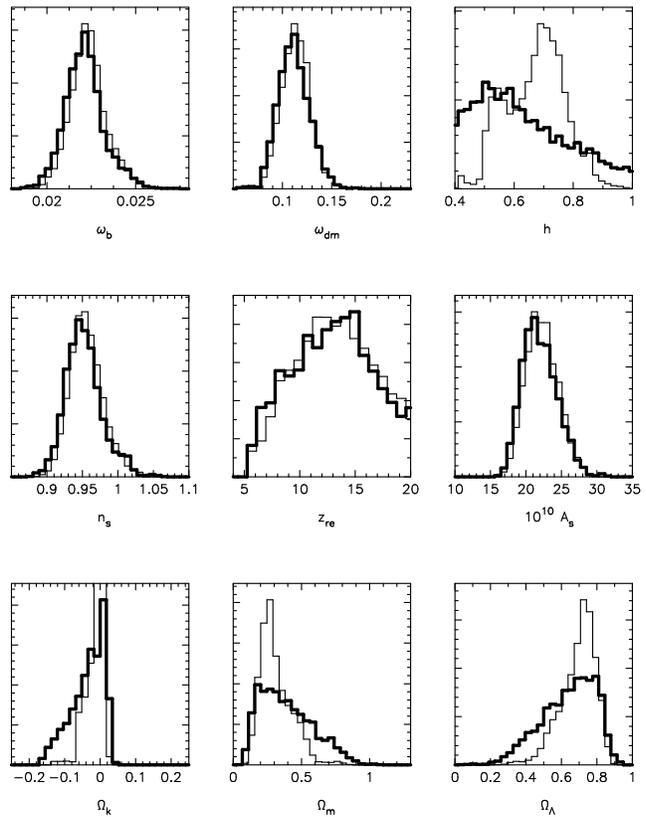}
\caption{The marginalised distributions for the cosmological
  parameters as inferred from data set 1 using the original
\cosmomc sampler (thin line) and the \cog sampler (thick line) with a
  chain length of 6500. \label{d1fig2}}
\end{figure}

The marginalised probability distributions of the cosmological
parameters obtained from the \cosmomc and \cog samplers are shown in
Fig.~\ref{d1fig2}.
We see that the distributions for the parameters $\omega_{\rm b}$,
$\omega_{\rm dm}$, $n_{\rm s}$, $z_{\rm re}$ and $A_{\rm s}$ are
consistent within the sampling uncertainties for the two samplers.
We note, however, that the marginalised distributions for $h$,
$\Omega_{\rm k}$, $\Omega_{\rm m}$ and $\Omega_\Lambda$ differ
  somewhat. In particular, we see that the distributions
produced by the \cosmomc sampler are significantly narrower than those
obtained using the \cog sampler. This provides a useful illustration
of the poorer mobility of the chain in the \cosmomc sampler along the
geometrical degeneracy. This leads to underestimation of the
confidence limits on the associated parameters. Indeed, in some cases, the
limits are underestimated by around a factor of two. 
Conversely, the explicit degeneracy direction engine in
the \cog sampler allows its chain to move freely 
around the parameter space. This ensures that the marginalised
distributions reflect the proper structure of the posterior.
It should be noted, however, that the underestimation of the
confidence limits by the \cosmomc sampler is usually not a serious
problem, since an informative prior on $h$ breaks the geometrical
degeneracy sufficiently that the resulting marginalised distributions
are reasonably accurate.

\subsection{Future CMB data}

The second data set (data set 2)
consists of simulated cosmic-variance limited measurements of the
CMB power spectrum at each multipole out to $\ell=2000$. Such data is
expected from the forthcoming Planck satellite mission. In detail, a
target concordance model was chosen with parameter values 
$\omega_{\rm b}=0.023$, $\omega_{\rm dm}=0.127$, $h=0.68$, $n _{\rm
  s}=1.0$, $z_{\rm re} = 16$, $A_{\rm s}=25$ and $\Omega_{\rm k}=0.0$.
The corresponding theoretical CMB power spectrum was calculated using
\camb with the flat-universe code switched off. In an earlier analysis
of data set 2, we noticed that discontinuities occurred in the
sample distribution which were associated with the switch over in the \camb
code from flat universes to universes with arbitrary curvature. This
inaccuracy in the \camb code is only important when analysing data of
such high precision, and has been previously identified (Lewis,
private communication). To create data set 2, we assumed that the
cosmic variance limit can be well-approximated by a Gaussian
distribution centred on the true $C_\ell$ value and having a
dispersion given by
\begin{equation}
\Delta C_\ell = \sqrt{\frac{2}{2\ell+1}}C_\ell.
\end{equation}
In practice, with real data, one should take into account that the probability
of any given $C_\ell$ value follows a $\chi^2$-distribution; this is
particularly important at low $\ell$ values.

\subsubsection{Acceptance rates and sampler performance}

\begin{figure}
\epsfig{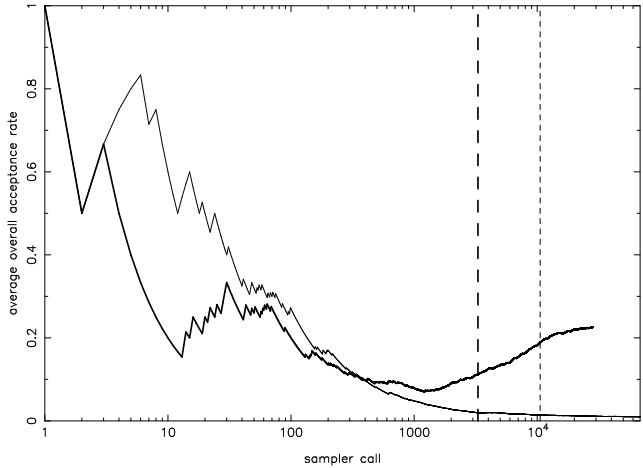}
\caption{As in Fig.~\ref{d1fig0}, but for data set 2. \label{d2fig0}}
\end{figure}

The average acceptance rates for the \cosmomc and \cog samplers
are plotted in Fig.~\ref{d2fig0} as a function of the number of calls
to the sampler.
Once again, the original \cosmomc sampler was given
the precomputed basic 6-parameter covariance matrix that comes
supplied with the \cosmomc software, which gives the \cosmomc sampler
some initial advantage. One sees, however, that the \cog sampler soon
overtakes and, in this case, there is a very large difference between
the equilibrium acceptance rates of the two samplers.
After many samples, the acceptance rate for the \cosmomc
sampler is found to be around 0.01. Even after 70000 sampler
calls for each chain, only around 600 post burn-in 
accepted samples were obtained for the best-performing chain. 
Conversely, the \cog sampler achieved
an equilibrium acceptance rate of 0.22, and required only 28000
sampler calls to obtain 6000 post burn-in 
accepted samples for each chain. We also note that, as a
result of the simulated annealing used in the \cog sampler, it
required only 3100 sampler calls were required to burn-in, as compared
with 10000 sampler calls for \cosmomc (in the latter case, the end of
burn-in was determined interactively by examining the posterior values
associated with the chain samples).

\begin{figure}
\epsfig{file=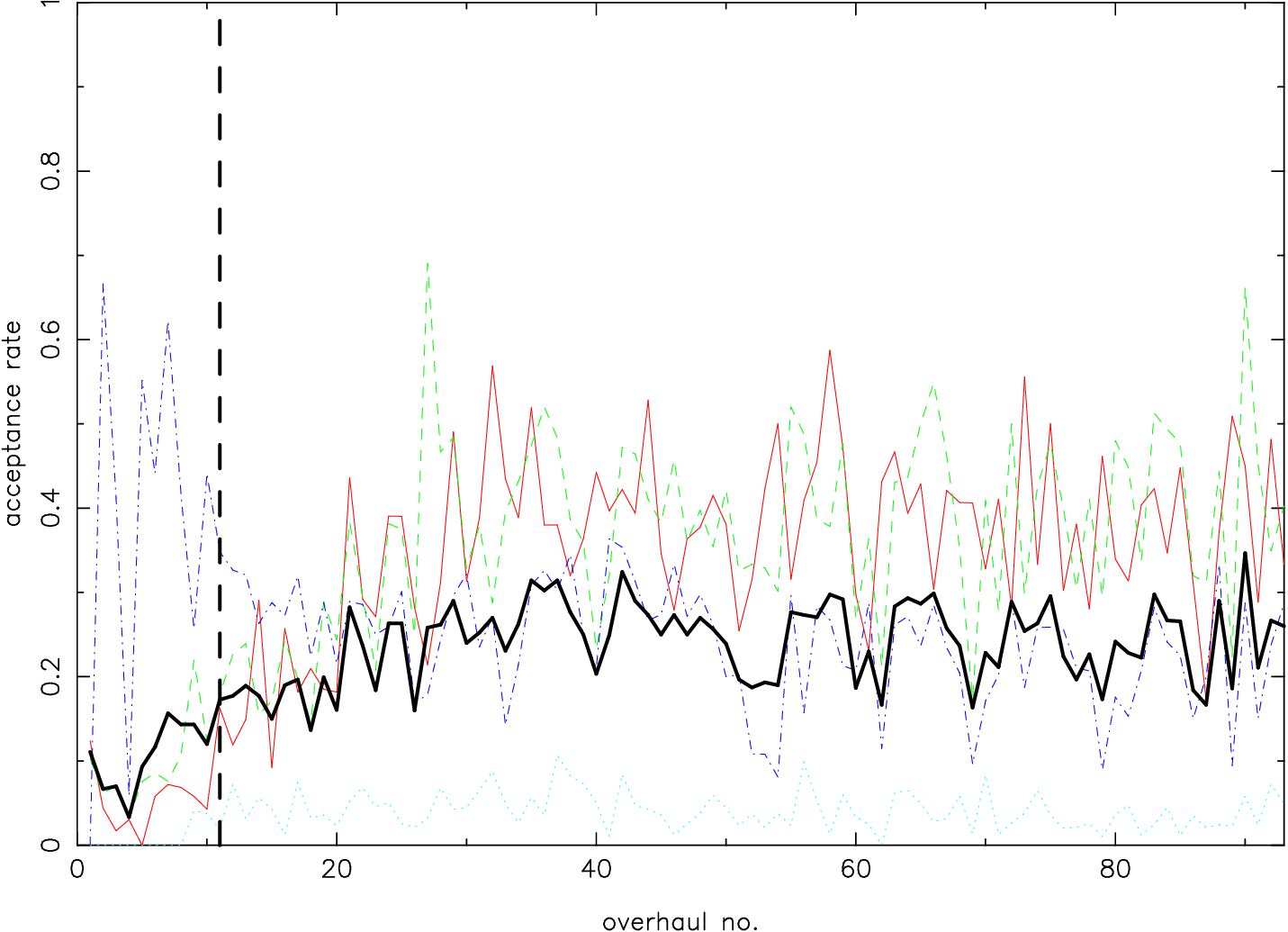,height=\linewidth, angle=-90}\\[10mm]
\epsfig{file=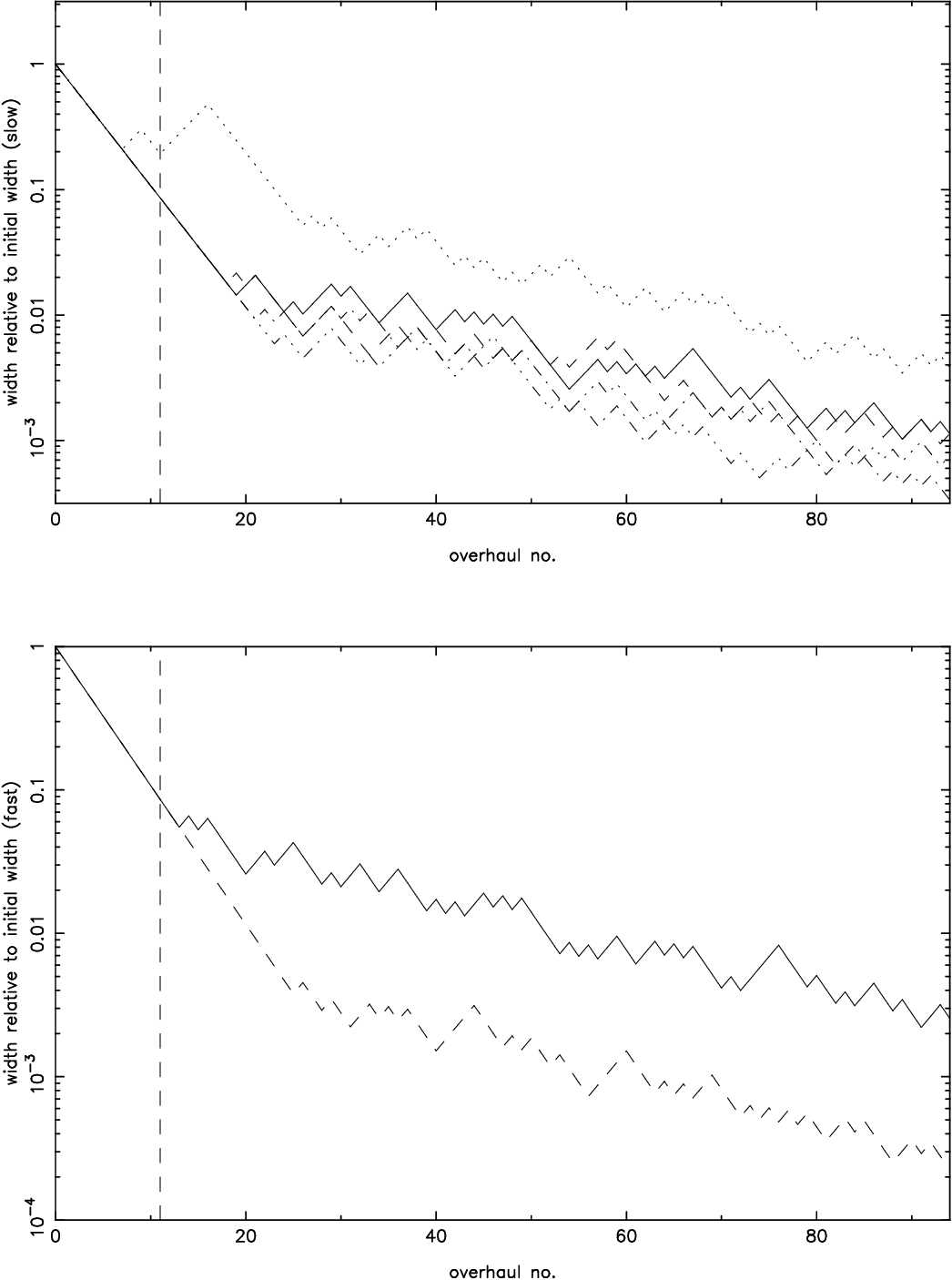,width=\linewidth}
\caption{As in Fig.~\ref{d1fig1}, but for data set 2. \label{d2fig1}}
\end{figure}

In Fig.~\ref{d2fig1} (top panel), 
we plot the acceptance rates for each of the four MCMC engines
used in the \cog sampler as a function of the number of overhauls.
Once again, by construction, the engines E1 and E2 have acceptance rates
of around 0.4 once equilibrium has been achieved. 
For E3 (principal direction changes), we again see that the initial
acceptance  rate is
zero, since there are not enough samples to compute the covariance
matrix, but in this case the acceptance rate remains low, reaching
an equilibrium value of only 0.05. We believe this behaviour is 
associated with the fact that, for data set 2, 
the corresponding multivariate Gaussian proposal distribution is a
poorer approximation to the true posterior than for data set 1.
For E4 (degeneracy direction changes), we see that
the inexactness in the degeneracies
and small numerical inaccuracies in both in \camb and the calculation
of the degeneracy parameters have a more profound effect for the
analysis of data set 2, resulting in a lower equilibrium acceptance
rate of around 0.25. As for data set 1, however,
successful proposals along the
degeneracy directions take the chain a considerable distance from its
original position and are thus very important.

\subsubsection{Inferred limits on cosmological parameters}

The marginalised probability distributions of the cosmological
parameters obtained from the \cog sampler are shown in
Fig.~\ref{d1fig2}.
\begin{figure}
\label{fig2}
\epsfig{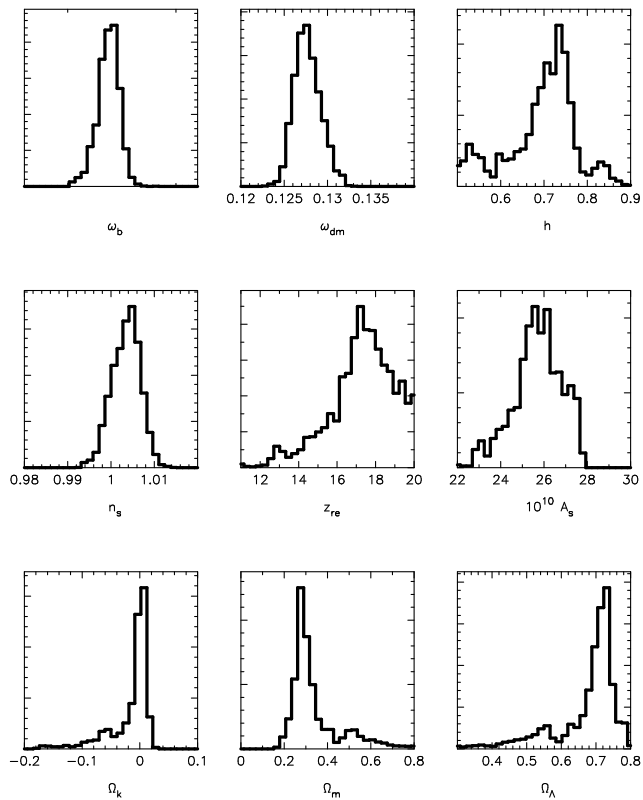}
\caption{As in Fig.~\ref{d1fig2}, but for data set 2 and using the
\cog sampler only. \label{d2fig2}}
\end{figure}
We do not plot the corresponding distributions for the \cosmomc
sampler, because it was only able to produce a total of 1500 
accepted post burn-in
samples from 70000 sampler calls for each of
the eight chains. This results in the corresponding marginalised
distributions being dominated by sampling error.

We note that the distributions for each parameter produced by the \cog
sampler contain the
corresponding true value at high probability. For
$\omega_{\rm b}$,
$\omega_{\rm dm}$, $n_{\rm s}$, $z_{\rm re}$ and $A_{\rm s}$ the
distributions are reasonably smooth. However, 
the marginalised distributions for $h$, 
$\Omega_{\rm k}$, $\Omega_{\rm m}$ and $\Omega_\Lambda$ contain some
additional features. In particular, we note that each of these 
distributions contains a `drop-out'; these occur at $h\approx 0.6$ and
$h \approx 0.8$, $\Omega_{\rm k} \approx -0.05$, $\Omega_{\rm m}
\approx 0.4$ and $\Omega_\Lambda \approx 0.6$. 
These features are, in fact, the projections of a single feature in
the multidimensional parameter space defining the 
geometrical degeneracy direction. We have separately examined the
marginalised distributions produced for each of the eight chains, and
found that all of them contain these features. It is therefore most
likely that this feature is an artifact in the \camb code. More importantly,
setting aside these drop-outs, we see that the geometrical degeneracy
is well-sampled, leading to wide tails on the marginalised
distributions for $\Omega_{\rm k}$, $\Omega_{\rm m}$ and
$\Omega_\Lambda$. This confirms the well-known result that, 
even with CMB data of Planck quality, 
the spatial  curvature of the universe cannot be well-constrained 
without the inclusion of informative priors.

\section{Discussion and conclusions}
\label{sec:conclusions}

We have presented a fast Markov-chain Monte Carlo sampler 
tailored to the problem of estimating cosmological parameters from 
measurements of the CMB total intensity power spectrum. 
This sampler employs a number of strategies to avoid the
difficulties encountered in the use of the standard multivariate
Gaussian proposal function used, for example, in the \cosmomc
software package (Lewis \& Bridle 2002). In particular, 
it achieves rapid convergence and produces reliable confidence limits
by using dynamic widths for proposal distributions, dynamic covariance
matrix sampling, and a dedicated proposal distribution for moving
along well-known degeneracy directions.
The new sampler module (called \cog) is trivially substituted for the 
existing sampler in the \cosmomc software.
In Section~\ref{sec:application}, we test the \cog sampler on the
current CMB data set and a simulated data set of the quality expected
from the Planck mission. In each case, the sampler produces
reliable marginalised distributions with considerably fewer
sampler calls than the native \cosmomc sampler. 

In future work, we intend to enhance the efficiency of the \cog 
sampler still further by allowing communication between chains.
At present running several independent chains on different processors
of a Beowulf cluster provides a useful means of diagnosing
convergence, but does not take advantage of the possibility of
sharing the information obtained by the chains on the shape of the
target density. In particular, we are investigating cross-chain
proposals based on genetic algorithms (see e.g. Davis 1991).

\section*{ACKNOWLEDGMENTS} 

We thank Charles McLachlan, Phil Marshall, John Skilling, Steve Gull and
Carolina \"Odman for many interesting conversations regarding
Markov-chain Monte Carlo sampling. We are also grateful to Antony
Lewis and Sarah Bridle for useful discussions and for making their
\cosmomc software publicly available.  AS acknowledges the support of
St. Johns College, Cambridge.

\bsp
\label{lastpage}
\end{document}